\documentclass[aps,prl,superscriptaddress,twocolumn,10pt,nobibnotes]{revtex4-2}

\usepackage{graphicx}
\usepackage{amssymb,amsmath,bbm,}
\usepackage[utf8]{inputenc}
\usepackage{comment}
\usepackage{braket}
\usepackage{enumerate}
\usepackage[normalem]{ulem}
\usepackage{hyperref}
\hypersetup{colorlinks=true,linkcolor=blue,citecolor=blue,urlcolor=blue}

\newcommand{\tr}[1]{\text{Tr} \left[ #1 \right]}
\newcommand{\Tr}{\mathrm{Tr}}
\newcommand{\ketbra}[2]{\ket{#1}\!\bra{#2}}

\usepackage[caption=false]{subfig}

\usepackage{soul}
\usepackage[table]{xcolor}
\usepackage{tcolorbox}
\definecolor{light_gray}{HTML}{f2f2f2}
\tcbset{fontupper=\normalsize,
colback=white, colframe=black, colbacktitle=light_gray,
coltitle= black, arc=0.5mm, center title}

\begin{document}

\title{Charging a quantum spin network towards Heisenberg-limited precision}

\author{Beatrice Donelli}
\affiliation{Istituto Nazionale di Ottica del Consiglio Nazionale delle Ricerche (CNR-INO), Largo Enrico Fermi 6, I-50125 Firenze, Italy}
\affiliation{European Laboratory for Non-linear Spectroscopy, Università di Firenze, I-50019 Sesto Fiorentino, Italy}

\author{Stefano Gherardini}
\email{stefano.gherardini@ino.cnr.it}
\affiliation{Istituto Nazionale di Ottica del Consiglio Nazionale delle Ricerche (CNR-INO), Largo Enrico Fermi 6, I-50125 Firenze, Italy}
\affiliation{European Laboratory for Non-linear Spectroscopy, Università di Firenze, I-50019 Sesto Fiorentino, Italy}

\author{Raffaele Marino}
\affiliation{Department of Physics and Astronomy, University of Florence, 50019 Sesto Fiorentino, Italy}

\author{Francesco Campaioli}
\email{francesco.campaioli@unipd.it}
\affiliation{Dipartimento di Fisica e Astronomia “G. Galilei” Università degli Studi di Padova, I-35131 Padua, Italy}
\affiliation{Padua Quantum Technologies
Research Center, Università degli Studi di Padova, Italy I-35131, Padova, Italy}
\affiliation{INFN, Sezione di Padova, via Marzolo 8, I-35131, Padova, Italy}

\author{Lorenzo Buffoni}
\affiliation{Department of Physics and Astronomy, University of Florence, 50019 Sesto Fiorentino, Italy}

\begin{abstract}
We present a cooperative protocol to charge quantum spin networks up to the highest-energy configuration, in terms of the network's magnetization. The charging protocol leverages spin-spin interactions and the crossing of a phase transition's critical point. Exploiting collective dynamics of the spin network, the cooperative protocol guarantees a precision advantage over any local charging protocol and leads to fluctuations (standard deviation) of the magnetization that scale as $1/N$, with $N$ being the number of spins in the network, i.e., the size of the spin battery. These findings mirror the Heisenberg limit for precision for parameter estimation in quantum metrology. We test our protocol on the D-Wave's Advantage quantum processing unit by charging sub-lattices with sizes ranging from $40$ to $5\,612$ spins, achieving the maximum magnetization and reaching a scalable charging precision beyond the standard quantum limit of $1/\sqrt{N}$. 
\end{abstract}

\date{\today}

\maketitle

The advancement of quantum technologies~\cite{Acin2018}, coupled with pressing global energy challenges~\cite{andrae2019prediction}, is sparking a surge in research at the intersection of quantum science and energy applications~\cite{Myers2022,Auffeves2022,metzler2023emergence}. This emerging area has unveiled compelling possibilities, such as energy advantages of quantum computing~\cite{Auffeves2022,Fellous-Asiani2023,Fellous-Asiani2023b}, work extraction from quantum coherences~\cite{Monsel2020,Francica2020,Shi2022,hernandez2022experimental,GherardiniTutorial}, optimal thermometry~\cite{Correa2015,Rubio2021,Hovhannisyan2021}, quantum heat engines~\cite{Campisi16NATCOMM7,vroylandt2018collective,Kloc19PRE100,GelbwaserKlimovsky19PRA99,Watanabe20PRL124,Latune20NJP8,Souza22PRE106,Kolisnyk23PRAPP19,Jaseem23PRA107}, heat transport~\cite{Pekola2015,Pekola2021}, power and precision advantages in many-body quantum batteries~\cite{Binder2015a,Campaioli2017,Andolina2019a,Friis2018,Santos2019a,Santos2020,Delmonte2021,Moraes2021,Abah2022,Imai2022,Dou2022a,Hu2022,Bakhshinezhad2023,mukherjee2021many}.

In particular, quantum batteries have garnered significant attention due to the super-extensive scaling of charging power~\cite{CampaioliRMP2024}---a phenomenon where charging speed increases with the battery size or capacity. For instance, arrays of $N$ quantum spins can achieve charging power densities that scale with $\sqrt{N}$~\cite{Binder2015a}. This scaling, linked to entanglement generation in unitary dynamics~\cite{Binder2015a,Campaioli2017}, has drawn analogies with the celebrated Grover's algorithm~\cite{giri2017review} and Heisenberg-limited parameter estimation~\cite{demkowicz2012elusive}, cornerstones of quantum computing~\cite{pittenger2012introduction} and sensing, respectively~\cite{GiovannettiNatPhot2011}. Similar enhancements are known~\cite{cong2016dicke}, or are now being actively explored, in other areas, such as light emission and absorption in atomic and molecular aggregates~\cite{higgins2014superabsorption}, energy and charge transport mediated by delocalisation~\cite{Taylor2018,Balzer2021}, heat currents in circuit quantum electrodynamics~\cite{Andolina2024_heat_current}, charging of quantum supercapacitors~\cite{Ferraro2019}, and energy fluctuations in quantum batteries~\cite{Friis2018,Santos2019a,Santos2020,Rosa2020}.

Superextensive scaling of power and precision in many-body quantum batteries hinge on interactions between their components~\cite{Campaioli2017,Le2018,Andolina2019a,Julia-Farre2020,Rosa2020,Gyhm22PRL128}, either direct~\cite{Binder2015a,Campaioli2017,Le2018,Gemme2023a} or mediated by an auxiliary charger subsystem~\cite{Andolina2018,Farina2019}. Although promising charger-mediated protocols have been identified~\cite{Ferraro18PRL120} and recently tested for quantum emitters coupled to optical cavities~\cite{Quach2022}, achieving superextensive charging power and precision in isolated spin systems remains challenging due to stringent interaction requirements~\cite{Gyhm22PRL128,Rossini2020, Rosa2020}. An intriguing avenue of exploration involves harnessing the nontrivial cooperative behaviours that emerge at quantum phase transitions~\cite{Wenniger2022,Buffoni2023cooperativequantum,buffoni2024collective}. However, as the field takes its first experimental steps~\cite{Quach2022,Joshi2022,Konopik2023,Wenniger2023}, a crucial question remains open: Can superextensive power and precision be realized on real-world quantum devices, such as noisy intermediate-scale quantum (NISQ) platforms~\cite{PRAYu2024} or other quantum platforms?

\begin{figure}[t]
    \centering
    \includegraphics[width=0.49\textwidth]{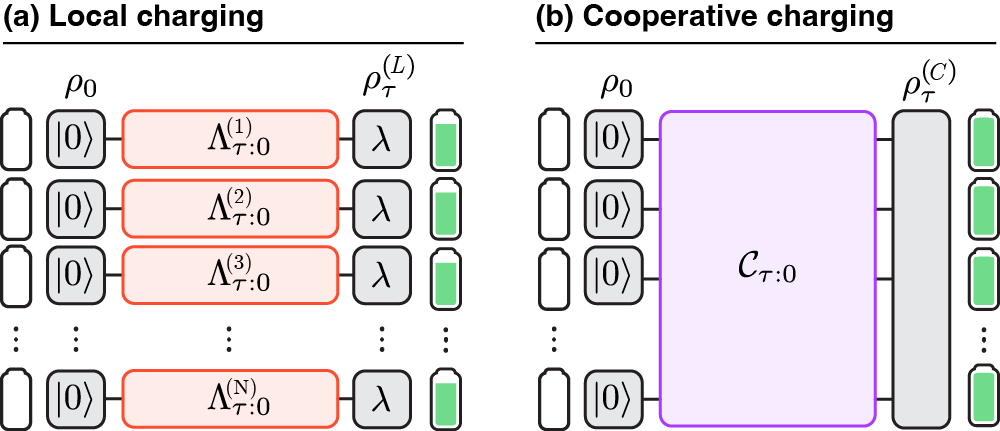}
    \caption{\textbf{Local and cooperative charging.}---We consider a many-body quantum battery given by $N$ spins charged via local fields and spin-spin interactions. The internal Hamiltonian $H_0$ is proportional to the network's magnetization $S_z = \sum_{k}\sigma_k^z$, with $\sigma_{k}^{z}$ denoting the local Pauli matrix $\sigma_z$ on the $k$-th spin, and the battery is initialized in the ground state of $H_0$. 
    (a) Local charging protocol, represented by dynamical maps $\Lambda_{\tau:0}^{(k)}$ acting on the $k$-th spin. The final state $\rho_\tau^{(L)} = \lambda^{\otimes N}$ is the product state of $N$ local mixed states $\lambda$.
    (b) Cooperative charging protocol, represented by the dynamical map $\mathcal{C}_{\tau:0}$ that charges the initial state $\rho_0$ to some state $\rho_\tau^{(C)}$. Here, we introduce a cooperative protocol that exhibits a $\sqrt{N}$ charging precision advantage over the local protocol.}
    \label{fig:local-cooperative}
\end{figure}

In this Letter, we introduce a cooperative protocol (CP) for the collective charging~\cite{RojoFrancs2024} of quantum spin networks that can reach Heisenberg-limited precision while relying on simple Ising spin-spin interactions~\cite{Mohseni2022} and the crossing of a phase transition's critical point~\cite{suzuki2012quantum}.
First, we show that the CP leads to fluctuations (standard deviation) in the final magnetization of an $N$-spin network that scale as $1/N$, 
in analogy with Heisenberg-limited quantum metrology~\cite{Giovannetti2004,GiovannettiNatPhot2011,Pezz2018}. Our CP offers a $\sqrt{N}$ precision advantage when compared with any local protocol (LP) whose precision follows the standard quantum limit (SQL), as illustrated in Fig.~\ref{fig:local-cooperative}. Then, we demonstrate the superextensive charging precision of the CP on a real-world device, i.e., the D-Wave's Advantage quantum annealer~\cite{dwdocs}, by charging networks of $40$ to $5\,612$ spins---the largest superconducting spin charging routine on a programmable platform realised to date. We conclude the Letter by discussing the conditions to achieve Heisenberg-limited charging precision and the application of the protocol in powering devices such as quantum computing platforms and single-atom engines.

\textit{Charging protocols.} 
We consider a battery comprised of $N$ uncoupled quantum spins. Its internal (non-interacting) Hamiltonian is $H_0 = \eta \, S_z$, where $\eta > 0$ is the local magnetization of the spin-lattice and $S_z := \sum_{k=1}^N\sigma_k^z$, with $\sigma_{k}^{z}$ denoting the Pauli matrix $\sigma_z$ on the $k$-th spin. For the sake of simplicity, $\eta\equiv 1$ henceforth.
The aim of a charging protocol is to inject as much energy as possible into the battery, which is initially assumed to be in the ground state $\rho_0$ of its internal Hamiltonian $H_0$, representing its lowest energy configuration. We denote $\ket{0}_k$ as the ground state of $\sigma_k^z$, such that $\rho_0 = \ketbra{\psi_0}{\psi_0}$ with $\ket{\psi_0}:=\bigotimes_k \ket{0}_k$.
A common strategy to inject energy consists in using unitary charging cycles, driven by some time-dependent Hamiltonian $H(t)$, which vanishes at the initial ($t=0$) and final ($t=\tau$) times of the protocol, i.e., $H(0)=H(\tau)=0$~\cite{CampaioliRMP2024}.

Here, we consider charging protocols governed by the time-dependent transverse field Ising Hamiltonian
\begin{equation}
\label{eq:H_s}
    H(t) = \mathcal{B}_x(t) S_x + \mathcal{B}_z(t) S_z + \mathcal{J}(t) \sum_{\langle j,k \rangle}^{N}\sigma_{j}^{z}\sigma_{k}^{z}\,,
\end{equation}
where $S_x := \sum_{k=1}^N\sigma_k^x$, $\langle j,k\rangle$ are the index pairs of coupled spins, and $\mathcal{B}_x(t), \mathcal{B}_z(t), \mathcal{J}(t)$ are the time-dependent `schedule functions' that determine the protocol. The choice of this Hamiltonian is motivated by the controllability of superconducting quantum annealers~\cite{hauke2020perspectives}, like D-Wave's Advantage QPU, and possible applications to Rydberg atom quantum simulators that operate in the ground-Rydberg basis~\cite{adams2019rydberg}. 
Since we are interested in the charging precision of real-world devices, we must consider the effect of relaxation and decoherence on the spin battery. To do so, we also generalize the charging protocols to allow for deviations from unitary, which we describe through completely positive and trace-preserving dynamical maps~\cite{Breuer2002}, as illustrated in Fig.~\ref{fig:local-cooperative}.

\begin{figure*}[ht]
    \centering
    \includegraphics[width =\textwidth]{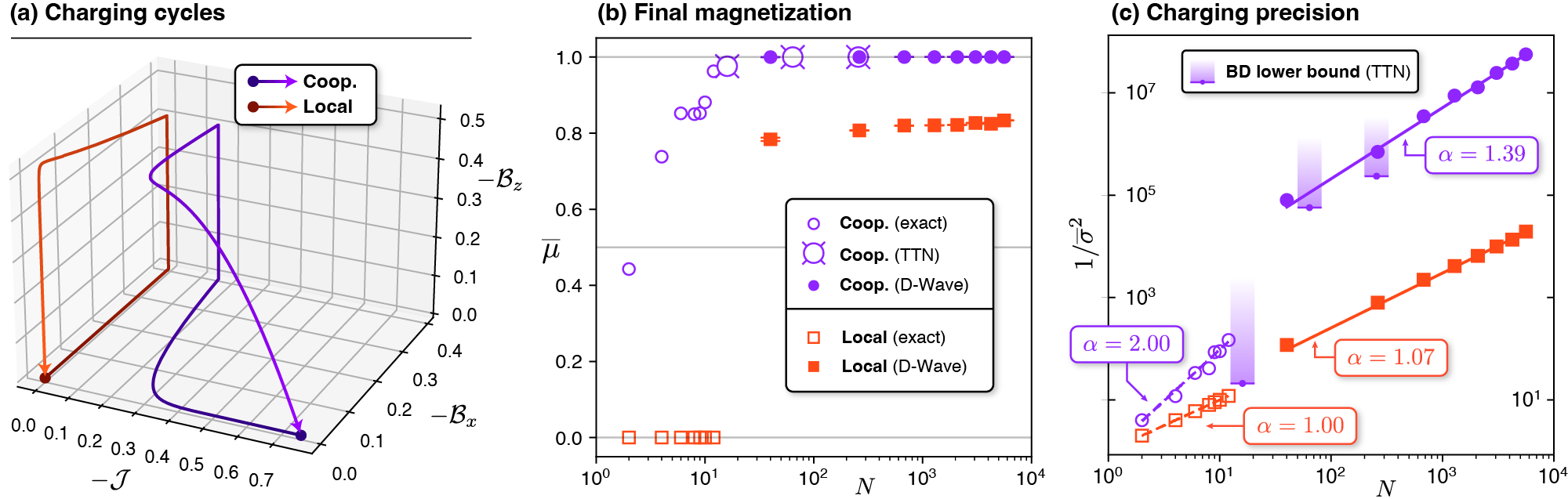}
    \caption{
    \textbf{Scaling of charging precision.}---{(a)} Local and cooperative charging protocols in the $(\mathcal{J},\mathcal{B}_x,\mathcal{B}_z)$ parameter space, expressed in GHz units. {(b)} Final magnetization of the spin network for the CP (purple markers) and LP (red markers). Numerical results shown with open markers; circles for exact simulations and spiked circles for tree tensor network (TTN). D-Wave results (solid markers) shown with error bars representing $\overline{\sigma}$. 
    Exact LP simulations have vanishing final magnetization. The discrepancy between numerical and measured data for the LP arises from unwanted cross-talk between the spins in D-Wave's Advantage QPU.
    {(c)} Final charging precision,  expressed as the inverse of $1/\sigma^2$ for CP (purple circles) and LP (orange squares). The lines are fit to the fluctuation model $f(x) = c N^\alpha$, with $\alpha \in [1,2]$; dashed for exact simulations and solid for D-Wave. TTNs calculations of the magnetization $\mu$ are used to lower bound the charging precision via the Bhatia-Davis inequality $\sigma\leq (1-\mu)(\mu+1)$~\cite{Bhatia2000}, here shown by lower-bounded purple gradients. Local charging precision follows the SQL ($\alpha \approx 1$), while cooperative charging leads to a precision advantage up to the HL ($\alpha = 2$). D-Wave QPU displays a scalable charging precision advantage with $\alpha = 1.39$.
    }
    \label{fig:results}
\end{figure*}

\textit{Charging via cooperative erasure protocol.} Our charging strategy is inspired by the cooperative erasure protocol~\cite{Buffoni2023cooperativequantum}, and it works as follows. Given a many-body battery operated by an uni-axial Ising ferromagnet within an environment at some finite temperature, we take a spin-spin interaction energy $\mathcal{J}$ such that the battery is in the ferromagnetic state, with no external magnetic field $\mathcal B_z$ and transverse field $\mathcal B_x$ applied. 
Starting from $\ket{\psi_0}$ we decrease the interaction energy $\mathcal{J}$ to a value that allows for the crossing of a critical point, where the battery gets into the paramagnetic state and becomes very sensitive to external fields. At the same time, we apply a transverse field $\mathcal{B}_x$ which helps in getting rid of eventual hysteresis effects and it has been shown to increase remarkably the performance of the protocol~\cite{buffoni2024collective,Buffoni2023cooperativequantum}.
Afterwards, by turning on the $\mathcal B_z$ field, the system aligns with its direction,  while the $\mathcal{J}$ and $\mathcal{B}_x$ are brought back to their initial values such that the protocol is cyclical and the system remains locked in a magnetized state due to the mechanism of spontaneous symmetry breaking (SSB)~\cite{goldenfeld2018lectures}. As a result, the spin ensemble accrues a positive net magnetization, which is equivalent to injecting energy into the spin battery. The CP is shown in Fig.~\ref{fig:results}~(a), where it is compared with a LP, obtained by setting $\mathcal{J}\to 0$; see End Matter (EM) and Supplemental Material (SM) for details.

\textit{Heisenberg-limited charging precision.} We now briefly discuss the scaling of energy fluctuations at the end of the charging protocols. Any local protocol drives the initial pure product state $\rho_0$ to a final product state $\rho_\tau^{(L)}$, which is generally mixed if decoherence and relaxation are considered. Assuming that all spins follow the same dynamics, the magnetization's fluctuations of $\rho_\tau^{(L)}$ read 
\begin{align}
    \label{eq:local_fluctuations}
    &\sigma_{(L)}^2 = \frac{1}{N}\big[1-(1-p)^2\cos^2 (2\theta)\big],
\end{align}
where $p\in[0,1]$ represents depolarisation from unit purity~\footnote{The purity of a state is $\mathcal{P} = {\rm Tr}\left[ \rho^2 \right]$.} of each spin, and $\theta\in[0,\pi/2]$ is a local rotation angle from the $z$-direction, i.e., the magnetization axis. Since $p$ and $\theta$ are independent of $N$, the energy fluctuations of any local charging protocol satisfy the standard quantum limit (SQL) $\sigma_{(L)} \propto 1/\sqrt{N}$.

Cooperative protocols, instead, can drive $\rho_0$ to highly correlated final states $\rho_\tau^{(C)}$, due to spin-spin interactions. We calculate the magnetization's fluctuations after a global rotation $\Theta\in[0,\pi/2]$ from the total magnetization axis, and a global depolarisation $P\in[0,1]$ from unit purity,
\begin{align}
    \label{eq:global_fluctuations}
    &\sigma_{(C)}^2 = \frac{P}{N} + (1-P)\big[1-(1-P)\cos^{2}(2\Theta)\big].
\end{align}
In general, the global parameters $\Theta$ and $P$ may depend on $N$. The Heisenberg limit (HL) $\sigma_{(C)} \propto 1/N$ can be obtained for $\Theta \propto 1/N$ and $P \propto 1/N^2$. See the EM for derivations and details.

We show that Heisenberg-limited charging precision can be achieved by simulating numerically ~\footnote{We simulate unitary charging cycles driven by $H(t)$ of Eq.~\eqref{eq:H_s} using exact propagation methods.} the charging protocols for Ising spin networks with nearest neighbour couplings and periodic boundary conditions for $N=2,4,6,8,9,10,12$. The results, shown in Fig.~\ref{fig:results}~(b) and~(c), indicate that the CP vastly outperforms the local one, leading to near-unit magnetization and fluctuations that scale as $1/N$. Note that when comparing the performance of different protocols, we ensure that their {\it energy cost} is the same~\cite{Binder2015a,Campaioli2017,Andolina19PRL122,Julia-Farre2020,CampaioliRMP2024}. To quantify the energy cost of a protocol we use the Hilbert-Schmidt (HS) norm $\|H\|_\mathrm{HS}:=\sqrt{\mathrm{Tr}[H^\dagger H]}$ of the Hamiltonian \eqref{eq:H_s}; see EM and SM for derivation and details.  

\textit{Demonstration on a quantum annealer.} The numerical results in Fig.~\ref{fig:results}~(c) suggest that the CP is promising towards Heisenberg-limited charging precision. However, how well does this approach translate to real quantum devices? Being quantum annealers, D-Wave QPUs are
not equivalent to a universal quantum computer and thus lack the control and stability necessary to lead to advantages based on entanglement as those that originate from the Greenberger-Horne-Zeilinger (GHZ) states in Heisenberg-limited parameter estimation~\cite{Pezze2018}. A hasty conclusion would be that charging precision beyond the SQL is unfeasible on D-Wave annealers. On the contrary, here we show that the Advantage QPU is sufficiently robust to achieve charging precision that scales far better than the SQL.

D-Wave's Advantage 6.4 quantum annealer consists in an ensemble of 5\,612 superconducting qubits that can be controlled via the time-dependent Hamiltonian \eqref{eq:H_s}, by mapping the schedule functions as follows:
\begin{equation}
    \label{eq:DWave-parameters}
    \mathcal B_x := -\frac{A(s)}{2}, \quad \mathcal B_z := g(t)h \frac{B(s)}{2}, \quad
   \mathcal J := J\frac{B(s)}{2},
\end{equation}
where $g(t)$ is a time-dependent parameter that, together with $h$, controls the strength of the local field along the $z$-direction, while $J$ controls the interaction strength between connected pairs of spins in the lattice. The time-dependent parameter $s$ in Eq.~\eqref{eq:DWave-parameters} determines the strength of the D-Wave's energy functions $A(s)$ and $B(s)$, expressed in GHz units (reduced Planck's constant $\hbar \equiv 1/(2\pi)$); see also EM and SM for details.

When implementing local and cooperative charging protocols on D-Wave's QPU, we set the total schedule time~\footnote{The choice of protocol duration was primarily informed by trial-and-error aimed at optimising the network magnetization at the end of the charging process.} to $\tau=60\:\mathrm{\mu s}$. For the CP, the bias term $h$ is set to $h_C = 0.5$ and the interaction term to $J=0.2$. In the LP, we set $J=0$ to turn off all spin-spin interactions, and carefully adjust the magnetic field parameter $h_L$ to ensure that the energy resources of each protocol are equivalent. The resulting expression for $h_L$ is
\begin{equation}\label{eq:rescaling_h_L}
    h_L = \sqrt{h_C^2 + \frac{n_C}{N} \frac{J^2}{g}} \,,
\end{equation}
where $n_C/N$ is the ratio of connectivity to the number of spins, and $g$ is the slope of $g(t)$ that we take as a linear ramp; see EM and SM for details. 

Fig.~\ref{fig:results}~(b) shows the final magnetization of the system as a function of $N$ for both protocols. The solid markers represent the mean value of the final magnetization $\overline{\mu}:= \sum_{i=1}^\nu \mu_i/ \nu$, obtained by averaging over $\nu=10^3$ repetitions of the magnetization measurements $\mu_i$. In the CP, $\overline{\mu}$ rapidly approaches 1, while in the local case the magnetization saturates at $\overline{\mu} \simeq 0.83$. These results correspond to a maximum charging power of $2.344 \times 10^{-16} \text{W}$ and $2.151 \times 10^{-16} \text{W}$ for the CP and LP respectively, as quantified in the SM.  

A substantial difference between the two protocols can be seen in the energy fluctuations, quantified by the standard deviation $\overline{\sigma}$ of the measured magnetization: 
\begin{equation}
\overline{\sigma}:= \frac{\sum_{i=1}^\nu (\mu_i-\overline{\mu})^2}{\nu-1}\,.
\end{equation}
The scaling of the charging precision as a function of $N$ is shown in Fig.~\ref{fig:results}~(c). For the LP, $1/\overline{\sigma}_L^2$ increases linearly with $N$, whereas for the CP, $1/\overline{\sigma}_C^2$ grows more rapidly.  The data of $1/\overline{\sigma}_L^2, 1/\overline{\sigma}_C^2$ are fitted using the function $f = c N^\alpha$, and we obtain the values shown in Tab.~\ref{tab:results} within the EM. We find that the CP is characterized by $\alpha = 1.39$, thus achieving superextensive charging precision, while the LP is limited to $\alpha = 1.07$.

\textit{Towards Heisenberg-limited charging precision.} In real quantum devices like D-Wave annealers, the spins are subject to energy relaxation and decoherence induced by the local environment and other sources of noise. Deviations from unitary charging tend to have a negative impact on the scaling of charging precision, preventing the saturation of the Heisenberg limit. To show it, we simulate the effect of decoherence, using a Lindblad master equation~\cite{Campaioli2024_tutorial} for $N=2,3,4,5$ spins. Our simulations are based on a simple model with a fixed decoherence rate $\gamma>0$ associated with local fluctuations of the transverse field; see SM for details. As the decoherence timescale $1/\gamma$ increases, the precision scaling exponent grows from $\alpha=1$ (SQL) to $\alpha=2$ (HL). With this model, we estimate the D-Wave's decoherence rate to be $\gamma\approx3.5\;\mathrm{kHz}$ ($\tau\approx 285\;\mathrm{\mu s}$) that is compatible with the estimated range reported in Ref.~\cite{imoto2023measurementenergyrelaxationtime}.

Another source of deviation from 
the ideal scenario is unwanted cross-talk, i.e., couplings, between nearby or even distant spins in the device's network. D-Wave annealers are well known to suffer from this problem~\cite{Zaborniak2021,Chern2023}, and some strategies can be used to mitigate---yet, not completely avoid---this effect, such as running a schedule on spins that are located apart from each other. To show the impact of cross-talk we simulate the LP by varying the strength of cross-talk couplings, here modelled with a constant and uniform interaction `bias' $J_\mathrm{CT}$ between spin pairs, where by definition $J_\mathrm{CT}=0$ in case of no cross-talks. In the absence of cross-talk, the LP leads to vanishing net magnetization, as shown in Fig.~\ref{fig:results}~(b). Instead, as $|J_\mathrm{CT}|$ grows, the charging precision exponent increases from $\alpha=1$ to $\alpha>1.15$ for $|J_\mathrm{CT}| > 0.002$. Based on this model, we estimate the strength of cross-talk between the spins of the Advantage QPU, obtaining $|J_\mathrm{CT}| \approx 1.3\cdot10^{-3}$ for tightly packed spins, and $|J_\mathrm{CT}|\leq 1.8\cdot10^{-4}$ for randomly chosen spins; see SM for details. Note that this cross-talk model differs from the random cross-talk model discussed in~\cite{dwcrosstalk}. 

These results suggest that slow decoherence and high spin connectivity are key to saturating the HL. Despite D-Wave Advantage being a noisy device, cooperative charging is sufficiently robust to noise and decoherence, leading to a performance that is well beyond the SQL.  

\textit{Conclusions.}
We demonstrate, theoretically and practically, that quantum Ising spin batteries can be charged cooperatively with a precision that scales superextensively, up to the Heisenberg limit. We conclude that noisy devices such as D-Wave's Advantage are robust enough to achieve these enhancements, by implementing the largest charging protocol of a superconducting spin battery to date. 

The cooperative protocol in this Letter could find application in powering quantum technologies with stringent precision requirements, such as quantum computing architectures based on Ising spin lattices~\cite{Menta2024} and reversible quantum operations~\cite{Chiribella2017,Chiribella2021}, as well as in exploring the thermodynamics of single-atom engines~\cite{ross2016singleatom}. Indeed, with the energy injected in the $5612$-spin network, one could power $\sim 10^5$ protocols of the single-atom heat engine in~\cite{ross2016singleatom} (see SM for details).

An interesting outlook could be to explore the compatibility of superextensive scaling in both charging power and precision. A trade-off between power and energy cost of charging can be inferred from fundamental relations for shortcuts to adiabadicity~\cite{Campbell2017}, suggesting that the two enhancements might be mutually exclusive. When exploring these aspects, we face the hurdle associated with the computational complexity of simulating the dynamics of large, complex, interacting spin networks at the crossing of critical points. For this reason, quantum devices like D-Wave annealers offer an interesting opportunity to study energy storage in many-body quantum systems.

\textit{Data availability.} The datasets and code generated and analyzed for this work are available upon reasonable request to the corresponding authors.

\begin{acknowledgments}
{\it Acknowledgements.} 
B.D.~and S.G.~acknowledges financial support the PNRR MUR project PE0000023-NQSTI funded by the European Union---Next Generation EU. S.G.~also acknowledges support from the PRIN project 2022FEXLYB Quantum Reservoir Computing (QuReCo). R.M. is supported by NEXTGENERATIONEU (NGEU) and funded by the Ministry of University and Research (MUR), National Recovery and Resilience Plan (NRRP), project MNESYS (PE0000006) – A Multiscale integrated approach to the study of the nervous system in health and disease (DN. 1553 11.10.2022). L.B. was funded by PNRR MUR Project No. SOE0000098-ThermoQT financed by the European Union--Next Generation EU.
\end{acknowledgments}

\bibliography{library}

\newpage
\clearpage
\section{End Matter}

\subsection{Protocol details}

D-Wave's Advantage QPU consists in the densely connected network of superconducting qubits known as Pegasus graph~\cite{dwdocs}, where each spin is coupled to around $15$ neighbors. Since the Pegasus graph is arranged in tiles, we can access subgraphs with a varying number of spins while keeping the Pegasus topology.

\begin{figure}[h!]
    \centering
    \includegraphics[width=0.8\columnwidth]{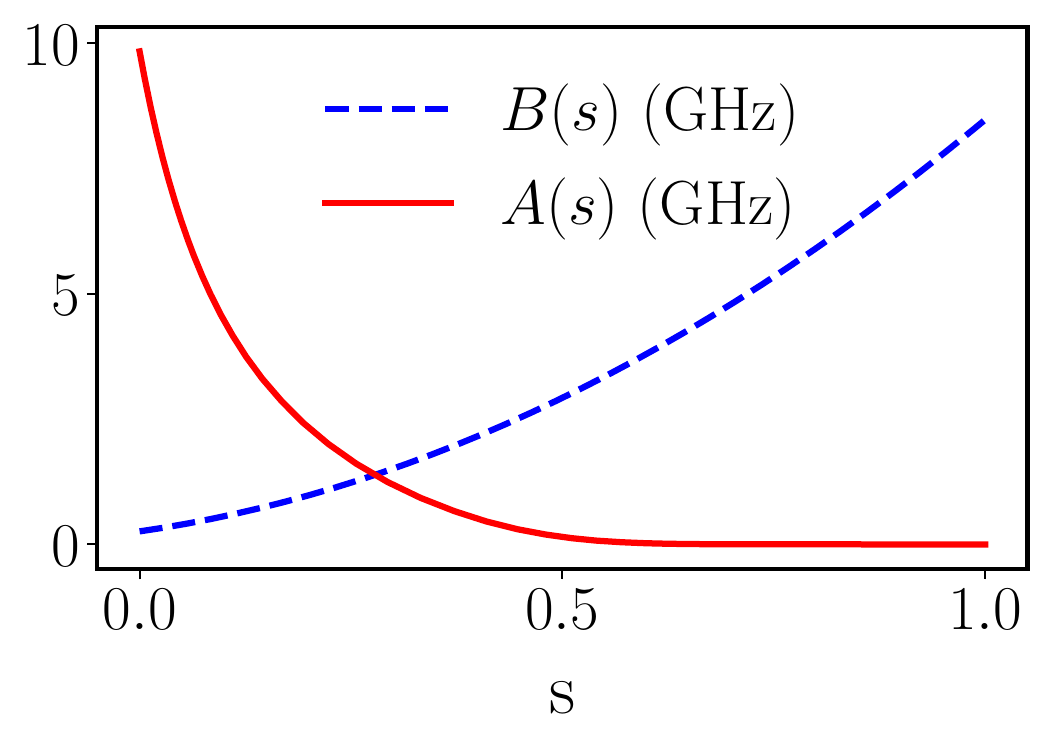}
    \caption{
    The energy factors $A(s)$ and $B(s)$ in units of GHz (i.e., Planck's constant equal to one). As one can see, in the region $s>0.5$ the transverse field is always turned off and the two functions cannot be independently controlled in time as they are both a function of $s$ \cite{dwdocs}.}
    \label{fig:ABfields}
\end{figure}

By inserting the correct terms in the Hamiltonian, as described in the main text, we obtain the following form
\begin{equation}
    H(t) = - \frac{ A(s) }{2}S_x + \frac{ B(s) }{2}\left[ g(t)h S_z + J \sum_{\langle j,k \rangle}^{N}\sigma_{j}^{z}\sigma_{k}^{z} \right] \,. 
\end{equation}
To carry out the desired charging protocol we use the schedule function $s(t)$ which depends on time and takes the form 
\begin{equation}
    s(t)=
    \displaystyle{
    \begin{cases}
    1 - 1.95\frac{t}{\tau} & \text{for} \quad 0 \leq t \leq \frac{\tau}{3}, \\
    0.35 & \text{for} \quad \frac{\tau}{3} \leq t \leq \frac{2\tau}{3}, \\
    -0.95 + 1.95\frac{t}{\tau} & \text{for} \quad \frac{2\tau}{3} \leq t \leq \tau \,,
    \end{cases}
    }
\end{equation}
where $\tau$ is the final time of the schedule.
The pre-factor energy functions
$A(s)$ and $B(s)$ of Eq.~\eqref{eq:H_s} are tabulated and provided by D-Wave, see Fig.~\ref{fig:ABfields}. The time-dependent function $g(t)$ is a further degree of freedom to tune the local magnetization over time.
In our protocols, $g(t)$ takes the following form:   
\begin{equation}\label{g_linear}
    g(t)=
    \displaystyle{
    \begin{cases}
    0 & \text{for} \quad 0 \leq t \leq \frac{\tau}{3}, \\
    -1 + 3\frac{t}{\tau} & \text{for} \quad \frac{\tau}{3} \leq t \leq \frac{2\tau}{3}, \\
    3 - 3\frac{t}{\tau} & \text{for} \quad \frac{2\tau}{3} \leq t \leq \tau \,.
    \end{cases}
    }
\end{equation}

\subsection{Details on charging precision}

\textit{Local charging precision.}---Without loss of generality, any local charging protocol drives an initial product state such as $\rho_0$ to a mixed product state,
\begin{equation}
    \label{eq:general_final_local}
\rho_\tau^{(L)}=\bigotimes_{k=1}^N\lambda_k,
\end{equation}
with
\begin{equation}
    \label{eq:local_state}
    \begin{split} \lambda_k &= 
    p\frac{\mathbbm{1}_k}{2} + (1-p)\big[\cos^2(\theta)\ketbra{1}{1}_k\\ &+ \sin^2(\theta)\ketbra{0}{0}_k + (1-\delta)\sin(\theta)\cos(\theta)\sigma_k^x\big],
    \end{split}
\end{equation}
where $p\in[0,1]$ represents depolarisation from unit purity, $\theta\in[0,\pi/2]$ is a rotation angle from the $z$-direction denoting the magnetization axis, and $\delta\in[0,1]$ represents dephasing onto the magnetization axis. Note that since the dynamics are completely local, the parameters $\delta$ and $\theta$ cannot scale with the number $N$ of spins involved.

The magnetization and fluctuations of such a state can be calculated analytically, note that the dephasing parameter $\delta$ does not affect either the magnetization or its fluctuations, since they are both quantities calculated by projecting the state onto the very same magnetization axis of the considered dephasing channel.

\textit{Cooperative charging precision.}---In general, cooperative charging protocols can drive any initial state to highly correlated states, due to the presence of spin-spin interactions. In particular, since the cooperative protocols drive the system through a critical point, correlations can in general grow and spread across the full system. While we do not have a general expression of the final state $\rho_\tau^{(C)}$ of the CP, its fluctuations are bounded by the Heisenberg limit (HL) as $\sigma_{(C)}^2 \propto 1/N^2$.

This can be seen analytically, by considering a cooperative final state given by a mixture of the highly-correlated GHZ state,
\begin{equation}
\ket{\mathrm{GHZ}_\Theta}:=\cos(\Theta)\ket{1}^{\otimes N}+\sin(\Theta)\ket{0}^{\otimes N},
\end{equation}
where $\Theta\in[0,\pi/2]$ is a global rotation, and the maximally mixed state, as follows
\begin{equation}
    \label{eq:correlated_approximation}
    \rho_\tau^{(C)} = (1-P)\ket{\mathrm{GHZ}_\Theta}\!\bra{\mathrm{GHZ}_\Theta}+P \bigotimes_{k=1}^N\frac{\mathbbm{1}_k}{2},
\end{equation}
where $P\in[0,1]$ is a global depolarisation parameter.
This leads to the following magnetization and fluctuations\begin{align}
    \label{eq:global_magnetization}
    &\mu_{(C)} = (1-P)\cos(2\Theta), \\
    &\sigma_{(C)}^2 = \frac{P}{N} + (1-P)\big[1-(1-P)\cos^{2}(2\Theta)\big].
\end{align}
Note that we have ignored the effect of dephasing on the magnetization axis since the considered observable are invariant to it.

\subsection{Details on field rescaling for resources equivalence}\label{HS_norm}

\begin{figure}[t!]
    \centering
    \includegraphics[width=0.9\columnwidth]{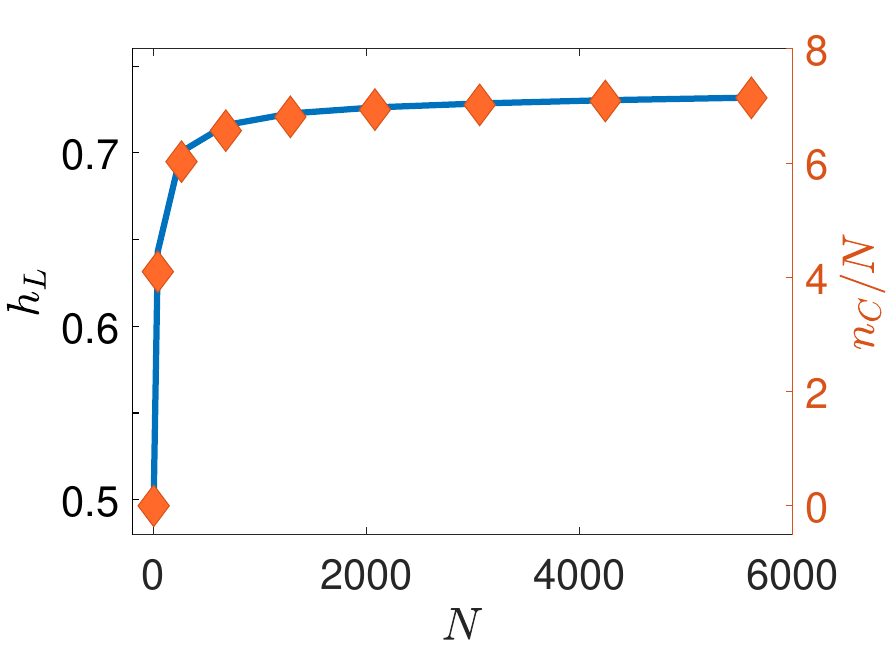}
    \caption{
    The values of $h_L$ as a function of $N$ are shown on the left axis (blue line), while the connectivity $n_C/N$ is shown on the right y-axis (orange diamonds).}
    \label{fig:hL_resources}
\end{figure}

In the LP, we rescale the external field $h_L$ to ensure that the resources used are equivalent to those in the CP. Specifically, we equate the HS norm of the cooperative and local Hamiltonians
\begin{equation}\label{Hc_Hl_norm}
    \|H_C\|_\mathrm{HS}=\|H_L\|_\mathrm{HS} \,.
\end{equation}
Using the norm expression of the driving Hamiltonian reported in the SM, the equality in \eqref{Hc_Hl_norm} between the norms in the two protocols expands to
\begin{equation}
\sqrt{2^N[N(\mathcal{B}_{x,C}^2+\mathcal{B}_{z,C}^2)+n_C\mathcal{J}^2]} = \sqrt{2^N N(\mathcal{B}_{x,L}^2+\mathcal{B}_{z,L}^2) } 
\end{equation}
where the subscripts $L$ and $C$ indicate the parameters for the LP and CP, respectively. Squaring both sides and substituting the expressions for the device's Hamiltonian coefficients as defined in \eqref{eq:DWave-parameters}, we obtain
\begin{align}\label{eq:rescaling_hL}
    \begin{split}
        \mathcal{B}_{x,C}^2 + \mathcal{B}_{z,C}^2 + \frac{n_C}{N} \mathcal{J}^2 = \mathcal{B}_{x,L}^2+\mathcal{B}_{z,L}^2 \\ 
        \frac{A(s)^2}{4} + \frac{B(s)^2 g(t)^2 h_C^2}{4} + \frac{n_C}{N} \frac{B(s)^2 J^2}{4} = \\ 
        = \frac{A(s)^2}{4} + \frac{B(s)^2 g(t)^2 h_L^2}{4} \,,    
\end{split}
\end{align}
which simplifies as 
\begin{equation}
    h_C^2 + \frac{n_C}{N} \frac{J^2}{g^2} = h_L^2 
\end{equation}
as given by \eqref{eq:rescaling_h_L} in the main text. In Fig.~\ref{fig:hL_resources} the values of $h_L$ (blue line, left y-axis), calculated from \eqref{eq:rescaling_hL}, are shown, along with the connectivity $n_C/N$ (orange diamonds, right y-axis) as functions of $N$. Using the parameters from the D-Wave runs $h_C=0.5$, $J=0.2$, $n_C/N \simeq 7$ and approximating the linear ramp as $g(t)\simeq 1$, we find $h_L \simeq \sqrt{0.5^2+7 \cdot 0.04} \simeq 0.73$, consistent with the values shown in Fig.~\ref{fig:hL_resources}.
These calculations confirm that the resources employed in both the CPs and LPs are effectively the same.

\subsection{Scaling of the charging precision}

In Tab.~\ref{tab:results}, we report the scaling of the charging precision, in the form of $1/\sigma^2$, that is inferred from the data by fitting to the model $f(x) = c N^\alpha$. Exact simulations of the charging schedule in the unitary limit display HL precision for the CPs ($\alpha = 2$), and SQL precision for the LPs ($\alpha = 1$). These results are calculated for 2D lattices with Torus topology, as detailed in the SM. The precision of local charging on D-Wave's Advantage QPU (Pegasus topology) is also perfectly compatible with the SQL ($\alpha =1.07$). Our cooperative charging schedule vastly outperforms the local one, despite using the same energetic resources, leading to a charging precision that scales beyond the SQL and towards the HL, with $\alpha = 1.39$.

\begin{table}[h!]
\begin{tcolorbox}[tabulars*={\renewcommand\arraystretch{1.2}}{@{\extracolsep{\fill}\hspace{2mm}}l@{\hspace{2mm}}|l|c|c},adjusted title=flush left,halign title=flush left, boxrule=0.5pt,title = {\hspace{-5pt}\textbf{Scaling of charging precision}}]
\textit{Protocol} & \textit{Topology}$\:$ & $c$ & $\alpha$ \\
\hline\hline
Local (exact) & Torus & 1.02 $\pm$ 0.00 & 1.00 $\pm$ 0.00 \\
\hline
Coop.~(exact) & Torus & 0.96 $\pm$ 0.44 & 2.00 $\pm$ 0.10 \\
\hline
Local (D-Wave) & Pegasus & 1.85 $\pm$ 1.09 & 1.07 $\pm$ 0.07 \\
\hline
Coop.~(D-Wave) & Pegasus & 350 $\pm$ 311 & 1.39 $\pm$ 0.11 \\
\hline
\end{tcolorbox}
\caption{
Scaling of the charging precision, fitted to the model $f(x)=c N^\alpha$.
}
\label{tab:results}
\end{table}

\clearpage
\newpage

\onecolumngrid
\section{Supplemental Material}
\twocolumngrid

\subsection{HS norm of local and driving Hamiltonian}

As discussed in the main text, we quantify the \textit{energy cost} of charging protocols by using the Hilbert-Schmidt (HS) norm $\|H\|_\mathrm{HS}:=\sqrt{\mathrm{Tr}[H^\dagger H]}$ of a charging Hamiltonian $H$. The HS norm is easy to calculate analytically and keeps track of every single component of the Hamiltonian that is turned on during the evolution, even if it does not affect it. 

We start by computing the HS norm in cases of interest for our purposes:
\begin{align}
    \label{eq:local}
    &\bigg\| \sum_{i=1}^N \sigma_i^{x}\bigg\|_\mathrm{HS} = \sqrt{N 2^N}, \\
    \label{eq:global_sx}
    &\bigg\| \bigotimes_{i=1}^N \sigma_i^{x}\bigg\|_\mathrm{HS} = \sqrt{2^N}, 
\end{align}
leading to up to $\sqrt{N}$ advantage for Eq.~\eqref{eq:global_sx} against Eq.~\eqref{eq:local}. The expression of Eqs.~\eqref{eq:local}-\eqref{eq:global_sx} comes from calculating the trace of the following Pauli matrices' combinations~\cite{borodulin2022core32compendiumrelations}:
\begin{align}
    &\tr{\sigma_i^{a}\sigma_j^{b}} = 2^N\delta_{ij}\delta_{ab}, \\
    &\tr{\sigma_i^{a}\sigma_j^{b}\sigma_k^{b}}=0, \\
    &\tr{\sigma_i^{a}\sigma_j^{a}\sigma_k^{a}\sigma_l^{a}} = 2^N(\delta_{ij}\delta_{kl} + \delta_{il}\delta_{jk}-\delta_{ik}\delta_{jl})\,,
\end{align}
with $a = x,y,z$.

Then, we consider a local Hamiltonian that, without loss of generality, can be expressed as
\begin{equation}\label{eq:H_loc}
    H_\mathrm{loc} := c_a \sum_{i=1}^N \sigma_i^{a} \,,
\end{equation}
where $\sigma_i^{a}$ is any element of a Lie Algebra for $\mathrm{SU(2)}$ (e.g., the Pauli matrices) such that $\Tr[\sigma_i^{a}\sigma_j^{a}] = 2^N\delta_{ij}$. The HS norm of the local Hamiltonian \eqref{eq:H_loc} is:
\begin{align}\label{eq:proof_local_HS}
    \|H_\mathrm{loc}\|_\mathrm{HS} &= \bigg\| c_a \sum_{i=1}^N \sigma_i^{a} \bigg\|_\mathrm{HS} = \sqrt{\tr{ c_a^2 \sum_{ij}\sigma_i^{a}\sigma_j^{a} }}=\nonumber\\
    &=c_a\sqrt{\sum_{ij}\tr{\sigma_i^{a}\sigma_j^{a}}}=
    c_a\sqrt{\sum_{ij}2^N\delta_{ij}}=\nonumber \\
    &=c_a\sqrt{\sum_{i}2^N} = c_a\sqrt{N 2^N}.
\end{align}

Similarly, we can compute the HS norm of the driving Hamiltonian of Eq.~\eqref{eq:H_s} in the main text that, for calculation purposes, we rewrite as 
\begin{equation}\label{eq:H_s_new_variables}
    H(t) = \mathcal{B}_x(t)\sum_{i=1}^N\sigma_i^{x}+\mathcal{B}_z(t)\sum_{i=1}^N\sigma_i^{z}+\mathcal{J}(t)\sum_{\langle i,j\rangle}\sigma_i^{z}\sigma_j^{z},
\end{equation}
where $\mathcal{B}_x(t) := -A(s(t))/2$, $\mathcal{B}_z(t) := g(t)h B(s(t))/2$, and $\mathcal{J}(t):= J B(s(t))/2$; see Eq.~\eqref{eq:DWave-parameters} in the main text. Moreover, we assume that $\sum_{\braket{j,k}}$ runs over a set of ordered, unique, coupling pairs $j,k$ with $n_C$ elements. 
In conclusion, using the formulas given above, the HS norm of \eqref{eq:H_s_new_variables} is 
\begin{equation}
\|H(t)\|_\mathrm{HS}=\sqrt{2^N[N(\mathcal{B}_x(t)^2+\mathcal{B}_z(t)^2)+n_{C}\mathcal{J}(t)^2]}\,.
\end{equation}

\subsection{Numerical simulations}

The numerical results included in Fig.~\ref{fig:results} have been obtained using exact propagation (open round markers) and tree tensor network methods (spiked open rounds markers). In both cases, we simulate local and cooperative charging protocols in the unitary limit, i.e., neglecting decoherence and relaxation. We use the same parameters considered in the D-Wave demonstration, fixing $J = -0.5$ and $h_C = -0.2$. 

\begin{figure}
    \centering
    \includegraphics[width = \columnwidth]{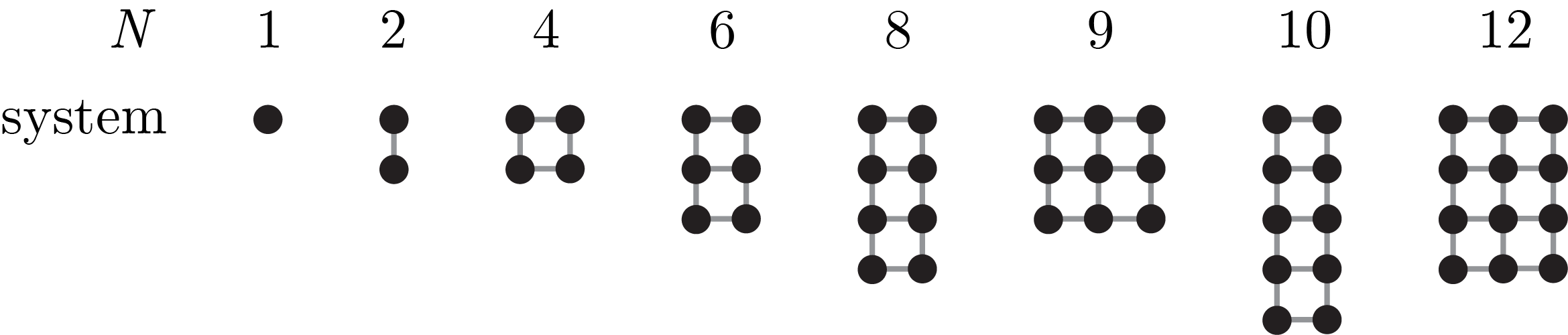}
    \caption{The spin lattices considered for the exact simulation.}
    \label{fig:systems}
\end{figure}

\begin{figure}
    \centering
    \includegraphics[width = \columnwidth]{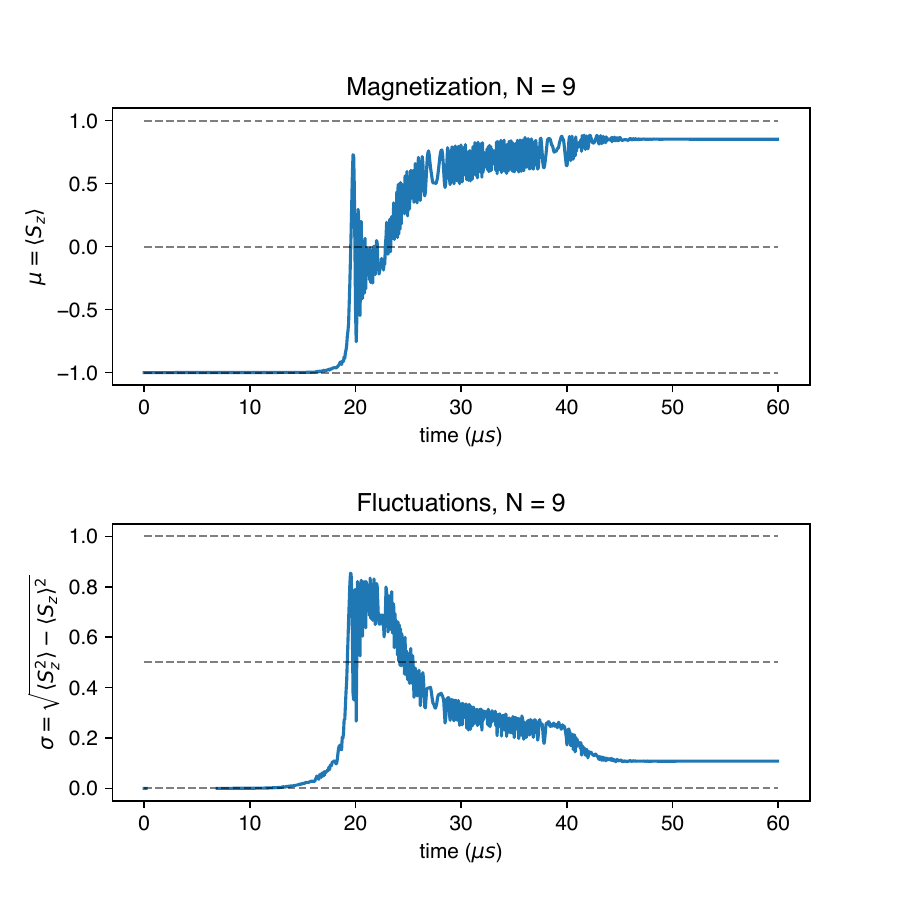}
    \caption{Example of exact propagation results for the magnetization $\mu$ and its fluctuations $\sigma$, of Eq.~\eqref{eq:observables_of_interest} for the 2D square lattice with periodic boundary conditions with $N=9$ spins.}
    \label{fig:exact}
\end{figure}

Exact propagation simulations have been performed using custom Python scripts based on sparse representations of states and operators and the action of the matrix exponential~\cite{Campaioli2024_tutorial}. To do so, we have used Scipy's structures \href{https://docs.scipy.org/doc/scipy/reference/generated/scipy.sparse.csr_matrix.html#csr-matrix}{\texttt{csr\_matrix}} and the method \href{https://docs.scipy.org/doc/scipy/reference/generated/scipy.sparse.linalg.expm_multiply.html#expm-multiply}{\texttt{expm\_multiply}}, respectively. Due to the exponential scaling of the Hilbert space associated with the system and explored during the dynamics, we simulated exactly the systems with $N\leq12$ shown in Fig.~\ref{fig:systems}. The simulations were run using a finite-difference time step $dt = 0.01$. Using this propagation method we could calculate exactly both the magnetization $\mu$ and its standard deviation $\sigma$, 
\begin{equation}\label{eq:observables_of_interest}
   \mu := \frac{\braket{S_z}}{N} = \frac{\tr{\rho \, S_z}}{N}\,, \quad\quad \sigma := \sqrt{ \frac{\braket{S^2_z} - \braket{S_z}^2}{N^2} }\,,
\end{equation}
where $\braket{S^2_z} := \tr{\rho \, S_z^2}$, and $\rho$ denotes a generic state of the spin battery as a whole.
An example of exact propagation results is shown in Fig.~\ref{fig:exact}.

Tensor network simulations have been performed using Quantum Tea Leaves version 1.1.8~\cite{qtealeaves}, an open-source Python package for tensor network simulation of many-body quantum systems. In this case, we have simulated the dynamics of 2D square lattices with periodic boundary conditions with a total number of spins $N=16,64,256$. The simulations where run using tree tensor network (TTN) ansatz, time-dependent variational principle propagation, using a finite-difference time step $dt = 0.01$ and a bond dimension $\chi =10$. Note that a finite bond dimension puts a limit on the amount of correlations that can be captured using a tensor network representation of the state of the system. Since the CP passes through a critical point, where correlations are expected to extend across the system, TTN simulations cannot faithfully capture the fluctuations $\sigma$. To circumvent this problem, we focus on the estimation of the magnetization $\mu$, which is a local observable and can be reasonably well approximated also using mean-field methods with minimal bond dimension $\chi = 1$. Then, we use the Bhatia-Davis inequality~\cite{Bhatia2000} to obtain a lower bound on the fluctuations of the magnetization, as discussed in the main text.

\subsection{Effect of decoherence}

To study the effect of decoherence we consider a simple time-dependent Lindblad master equation~\cite{Campaioli2024_tutorial} to the dynamics of the density operator $\rho_t$ representing the state of the system,
\begin{equation}
\label{eq:lindblad}
    \dot{\rho}_t = -i[H(t),\rho_t]+\gamma\sum_{k=1}^N \bigg(L_k\rho(t)L_k-\frac{1}{2}\bigg\{L_k^\dagger L_k^{\phantom{\dagger}},\rho_t\bigg\}\bigg),
\end{equation}
where $[,]$ is the commutator, $\{\cdot,\cdot\}$ is the anticommutator,
where $H(t)$ is the driving Hamiltonian, and $L_k = \bigotimes_{j=1}^N[\sigma_k^x\delta_{j,k}+(1-\delta_{j,k})\mathbbm{1}_k]$ are time-independent local coupling operators associated with $\mathcal{B}_x(t)$ noise. 

\begin{figure}
    \centering
    \includegraphics[width = \columnwidth]{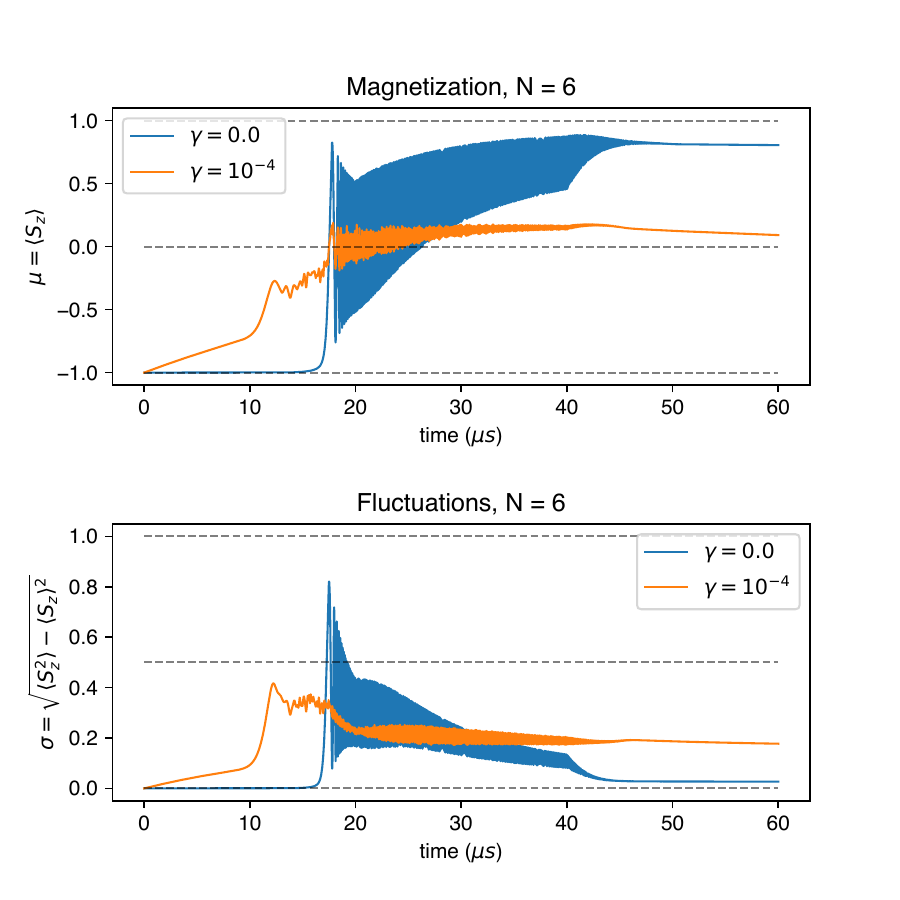}
    \caption{Example of exact propagation results for the unitary ($\gamma=0.0$) open ($\gamma =10^{-4}$) dynamics of the system with $N=6$. The local Lindblad operators $\sigma_k^x$ limit the efficacy of the charging protocol and increase the fluctuations in the magnetization.}
    \label{fig:open_exact}
\end{figure}

The Lindblad master equation is implemented using the Python package QuTiP's method \href{https://qutip.readthedocs.io/en/latest/guide/dynamics/dynamics-data.html}{\texttt{mesolve}}, which is exact, by varying the strength of the decoherence rate $\gamma$. For $\gamma=0$ we recover Unitary dynamics, while as $\gamma$ grows we increase the strength of decoherence, which leads to a slower scaling of the charging precision, as discussed in the main text. An example of exact propagation results for open dynamics is shown in Fig.~\ref{fig:open_exact}. 

\begin{figure}
    \centering
    \includegraphics[width=0.48\textwidth]{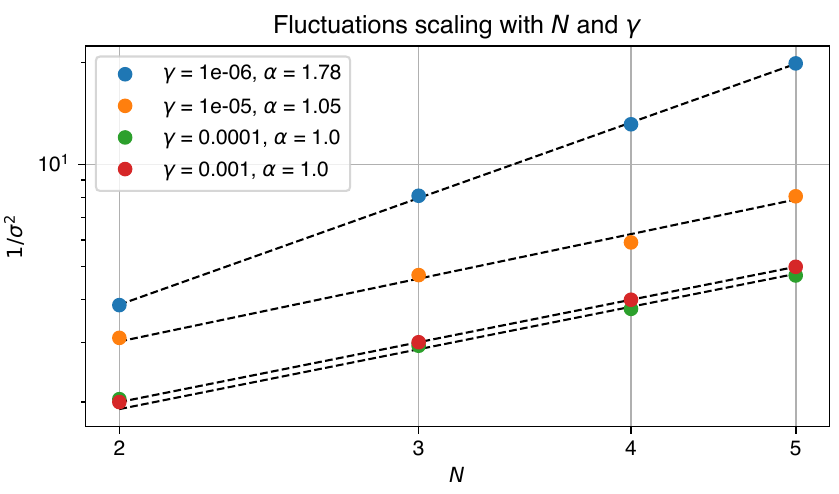}
    \caption{Scaling of magnetization fluctuations as a function of decoherence rate $\gamma$. The data points are obtained using the time-dependent Lindblad master equation of Eq.~\eqref{eq:lindblad} for the CP and fitted to the model $f = c N^\alpha$, to extract the scaling exponent. For $\gamma=0$ we recover $\alpha = 2$.}
    \label{fig:open_fits}
\end{figure}

By fitting the scaling of charging precision and comparing it to the one obtained from the D-Wave charging protocol we are able to estimate the decoherence rate of D-Wave. The computational complexity of these simulations scales with $2^{2N}$ since we have to propagate the density operator $\rho_t$ instead of the state vector $\ket{\psi}_t$, therefore, we only simulate systems with $N=2,3,4,5$. The results for the scaling of fluctuations are shown in Fig.~\ref{fig:open_fits} 

\subsection{Effect of cross-talk}

With cross-talk, we refer to unwanted ``residual'' couplings between sites in the D-Wave Advantage spin network. These might have various origins and are known to also depend on the choice of spins used during a schedule. For example, running the same protocol on closely packed spins or distant ones can lead to different outcomes. 

\begin{figure}
    \centering
    \includegraphics[width=0.48\textwidth]{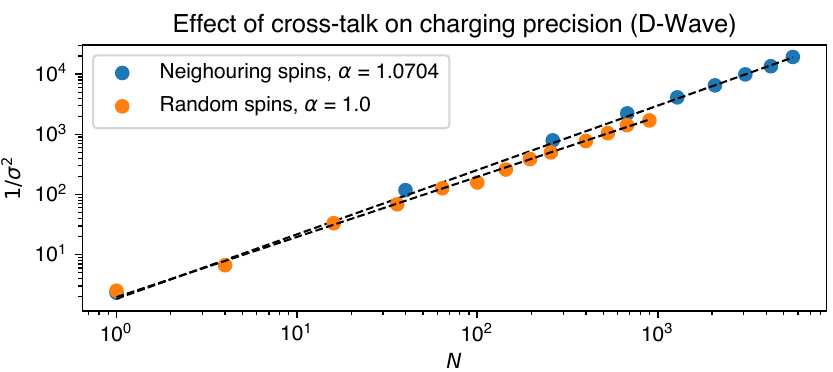}
    \caption{Effect of cross-talk on D-Wave charging precision, studied by simulating the LP for closely packed (neighbouring) spins (blue round markers) and randomly chosen spins across the full graph (orange round markers). To avoid size effects we limit random spin arrangements to $N<1000$. By fitting the scaling exponent of the inverse fluctuations we observe that charging precision is effectively linear in both cases but lower for random spins. We attribute that to the presence of weak but non-negligible cross-talk in the case of closely packed spins.}
    \label{fig:dwave_cross_talk}
\end{figure}

To study the effect of cross-talk we run the LP on sub-lattices of adjacent spins and on randomly chosen spins across the full graph of D-Wave Advantage QPU. The magnetization and fluctuations, shown in Fig.~\ref{fig:dwave_cross_talk}, indicate that charging precision scales as in the SQL in both cases, but with a slower slope in the randomly scattered spins. We interpret this as a results of lower cross-talk between spins that are further apart from each other. 

\begin{figure}[t]
    \centering
    \includegraphics[width=0.48\textwidth]{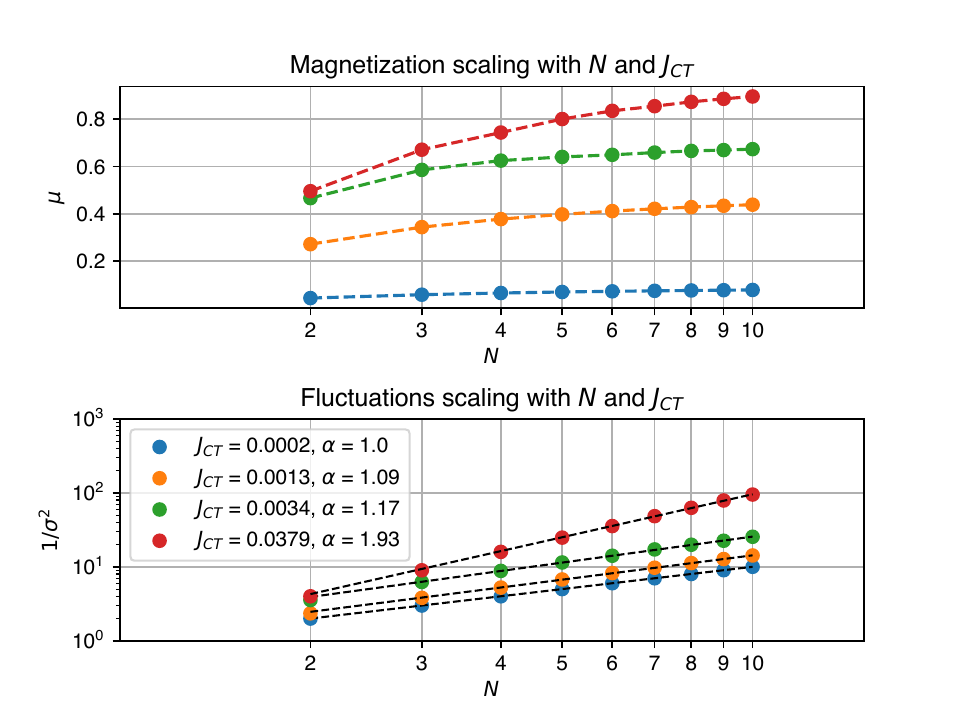}
    \caption{Scaling of the final magnetization and its fluctuations as a function of decoherence rate $\gamma$. The data points are obtained from exact propagation methods for the unitary CP. The fluctuations data points and fitted to the model $f = c N^\alpha$, to extract the scaling exponent. For $J_{CT}=0$ we recover local charging scaling $\alpha = 1.0$, while as $J_{CT}$ grows also $\alpha$ increases.}
    \label{fig:cross_talk}
\end{figure}

To simulate the effect of cross-talk we run exact numerical simulations of the unitary charging protocol by introducing a small uniform bias in the spin-spin couplings $J_{CT}$. The results, shown in Fig.~\ref{fig:cross_talk} are used to estimate the effect of cross-talk on the LPs. Even weak cross-talk $J_{CT}\approx 10^{-3}$, triggers significant deviations in the magnetization scaling, with respect to local charging. Furthermore, cross-talk also increases the charging precision exponent $\alpha$. We use this approach to estimate the strength of cross-talk for tightly packed spins and randomly chosen spins. As expected, randomly chosen spins appear to be affected by a weaker amount of cross-talk, leading to slower scaling of charging precision.

\subsection{Amount of Watts stored in a D-Wave's Advantage QPU with Pegasus topology}

To study the maximum power $P$ stored using our cooperative approach, we consider $N=5612$ qubits in a D-Wave's Advantage QPU with Pegasus topology, operating over a time window of $\tau=60 \mu \text{s}$. 

Given that the initial energy of the spin battery is $E_{0} = -\frac{B(s=1)}{2}N$ (with all spins initially down), we compute the power $P$ as $P = \frac{E_{\tau} - E_{0}}{\tau}$, where $E_{\tau}$ represents the final energy of the battery after the charging protocol has been applied. $E_{\tau}$ is determined by the number of spins that flip relative to the initial state. Therefore, $E_{\tau}=(1-f)E_{0}$, where $f$ is the fraction of flipped spins, and $f=\frac{\mu +1}{2}$, with $\mu$ being the magnetization, $\mu \in [-1, 1]$. From Fig.~\ref{fig:results} panel (b), we obtain $\mu_{C}=0.999$ and $\mu_{L}=0.833$, for the CPs and LPs, respectively. Using the value $B(s=1)=7.57 \text{ GHz}=5.016\times 10^{-24}\text{ J}$, as specified in the documentation of D-Wave's Advantage QPU with Pegasus topology, we determine the values of power $P_C=2.344 \times 10^{-16} \text{W}$ and $P_L=2.151 \times 10^{-16} \text{W}$ for the CPs and LPs, respectively.

In \cite{ross2016singleatom}, the authors reported the realization of a classical single-atom heat engine, where the working agent is an ion held within a modified linear Paul trap. From direct measurements of the ion dynamics, they determined the thermodynamic protocols for various temperature differences in the reservoirs. Using these protocols, they evaluated the engine's power $P_{\rm engine}$, taking the value $P_{\rm engine} = 3.4\times 10^{-22} \text{W}$ when the temperature differences of the reservoirs are set at $60\mathrm{mK}$. 

Therefore, a quantum spin battery as the one shown in the paper could store the energy equivalent to $6.894 \times 10^5$ engines using the CP and $6.327 \times 10^5$ engines using the LP, meaning that the CP outperforms the LP by a factor of $5.670 \times 10^4$. 

\end{document}